\documentclass{ws-p8-50x6-00}

\def\Journal#1#2#3#4{{#1} {\bf #2}, #3 (#4)}

\def\PREP{\em Phys. Rep.}

\def\PRC{{\em Phys. Rev.} C}

\def\EPJA{{\em Euro. Phys. J.} A}
\def\ANNP{\em Ann. Phys. (N.Y.)}
\def\PPNP{\em Prog. Part. Nucl. Phys.}
\def\AnRev{\em Ann. Rev. Phys. Sci. (N.Y.)}
\def\AOP{\em Adv. Nucl. Phys. (N.Y.)}
\def\JPG{{\em J. Phys.} G}

\begin{document}

\title{Many-body Theory at Extreme Isospin}

\author{H. Lenske, C.M. Keil and F. Hofmann}
\address{Institut f\"ur Theoretische Physik, Universit\"at
Giessen,
Heinrich-Buff-Ring 16, D-35392 Giessen, Germany\\
E-mail: horst.lenske@theo.physik.uni-giessen.de
 }


\maketitle

\abstracts{ The structure of nuclei far off $\beta$-stability is
investigated by nuclear many-body theory. In-medium interactions
for asymmetric nuclear matter are obtained by (Dirac-) Brueckner
theory thus establishing the link of nuclear forces to free space
interactions. HFB and RPA theory is used to describe ground and
excited states of nuclei from light to heavy masses. In extreme
dripline systems pairing and core polarization are found to be
most important for the binding, especially of halo nuclei. The
calculations show that far off stability mean-field dynamics is
gradually replaced by dynamical correlations, giving rise to the
dissolution of shell structures.}

\section{Introduction}\label{sec:intro}

Nuclear many-body theory provides a very general scheme allowing
to investigate the whole variety of nuclear systems from (purely
theoretical) infinite nuclear matter and macroscopic objects as
neutron star matter down to the microscopic level of finite
nuclei. Testing the concepts and methods of nuclear many-body
theory at the extremes as found in exotic nuclei is an important
and challenging question for structure physics. The vast amount of
new data on nuclei far off stability \cite{ov} has initiated
corresponding efforts on the theoretical side, ranging from new
impacts on cluster models and {\it ab initio} shell model
calculations for light nuclei to mean-field calculations over the
entire mass range. New phenomena, like halo and skin formation,
have been explained or predicted by nuclear theory, documenting
the progress made over the last years.

Until very recently, data were mainly available for ground state
properties of exotic nuclei. At present, the situation is changing
because the progress in experimental techniques allows to obtain
information also on dynamical processes in neutron- and
proton-rich nuclei. For theory this means to extend the methods to
a new sector of phenomena, e.g. inelastic and charge exchange
response functions and the fragmentation patterns of single
particle strength functions in breakup and transfer reactions. The
description of such phenomena clearly requires new approaches,
accounting properly for new effects as the coupling to the nearby
continuum and extreme isospin. Since one is entering unexplored
territories models providing intrinsically an extrapolation scheme
for interactions and the resulting nuclear dynamics are requested.
Phenomenological approaches like Skyrme theory and relativistic
mean-field theory are moderately successful in this respect, but
their flexibility and generality is constraint by the assumed
operator structure of the model Hamiltonian or Lagrangian and the
ability to determine the model parameters by fits to data only.

An important alternative is to approach the problem
microscopically. A clear advantage of such a program is that - at
least in principle - investigations can be based on a systematic
order-by-order hierarchy of interaction diagrams, as typical for
many-body theory, thus avoiding ambiguities and taking advantage
of results being tested independently in other regions of nuclear
physics. Taking this as a guideline the approach presented in this
contribution starts from calculating interactions in asymmetric
nuclear matter by Brueckner theory. As discussed in section
\ref{sec:anm} applications to the equation of state of infinite
matter, neutron star matter and neutron star structure
calculations give information on the global properties of the
interaction model, e.g. indicating the importance for going beyond
the pure ladder approximation inherent to Brueckner theory.

Applications of the microscopic approach to finite nuclei are
discussed in section \ref{sec:pair}, illustrating pairing in
dripline nuclei for $^{11}$Li, and core polarization far off
stability is the subject of section \ref{sec:dcp} as examples for
the transition into a new dynamical regime of correlation
dynamics. A summary and an outlook are given in section
\ref{sec:summary}.

\section{Interactions in Asymmetric Nuclear Matter}\label{sec:anm}

\begin{figure}[h]
\begin{center}
\epsfxsize=15pc 
\epsfbox{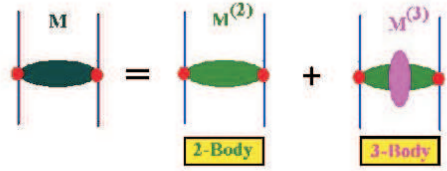} 
\caption{The in-medium NN scattering amplitude including 2- and
3-body contributions, respectively. The 3-body pieces, being of
order $\sim \cal{O}(\rho)$, result from the coupling to nucleon
resonances and polarization graphs. \label{fig:fig1}}
\end{center}
\end{figure}

Good confidence on the free NN-interaction has been obtained with
field theoretical meson-exchange models like the Bonn potentials
\cite{Machleidt}, showing that the ladder approximation is an
appropriate scheme in free space. In a medium, however, the
situation is less certain in the sense that additional classes of
diagrams will contribute, as illustrated in Fig.\ref{fig:fig1}.
The in-medium pieces introduce an additional density dependence
and contribute new types of operators, not found in free space
interactions \cite{Urbana,Coon}. Among those, the coupling of
mesons to medium polarization modes and three body forces from
intermediate excitations of nucleonic resonances (TBF) will alter
also the isospin structure of interactions. Little is known about
the dependence of these contributions on varying the
proton-to-neutron composition.

In ref.\cite{dejong} Dirac-Brueckner Hartree-Fock (DBHF) theory
was used to investigate 2-body interactions in infinite matter
over a large range of asymmetries. In-medium meson-nucleon
coupling constants were extracted for the isoscalar $\sigma$ and
$\omega$ and the isovector $\rho$ and $\delta$ mesons,
respectively \cite{dejong}. In all meson channels a pronounced
dependence on the isoscalar bulk density is found while the
dependence on the asymmetry is close to negligible. Hence, to a
very good approximation in-medium strong interactions remain
intrinsically independent of isospin thus conserving the
fundamental isospin symmetry. Using these coupling constants in
the density dependent relativistic hadron (DDRH) field theory
\cite{DDRH}  finite nuclei \cite{HKL}, hypernuclei \cite{hyper}
and neutron stars \cite{ns} are well described, thus underlining
the success of a microscopic approach.

\begin{figure}[t]
\begin{center}
\begin{minipage}[t]{12.2 cm}
\epsfig{file=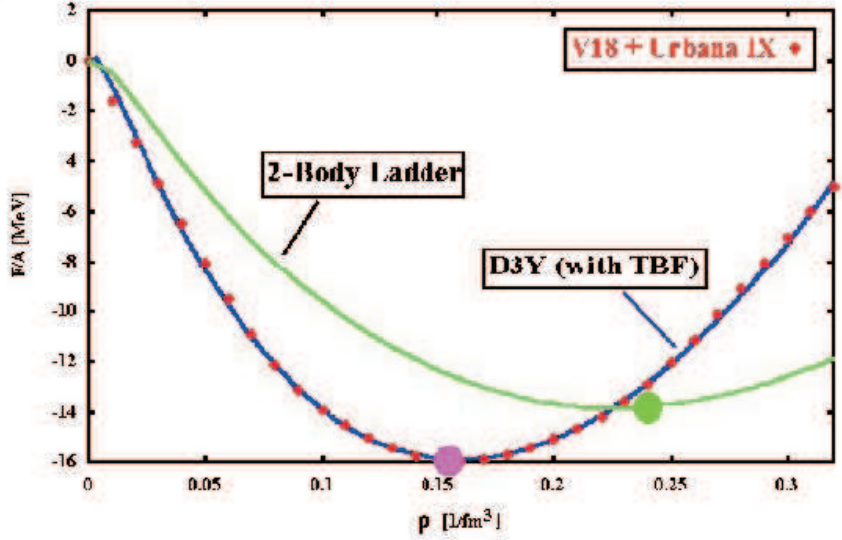,width=11.5 cm}
\end{minipage}
\begin{minipage}[t]{12 cm}
\caption{Equation of state of symmetric nuclear matter in
non-relativistic Brueckner theory without and with 3-body
interactions. The saturation point are indicated by full circles.
For comparison, variational result (dots) of the Urbana group
\protect\cite{Urbana} are also shown. \label{fig:fig2}}
\end{minipage}
\end{center}
\end{figure}

In Fig.\ref{fig:fig2} the equation of state obtained in
non-relativistic theory is shown, including results in ladder
approximation only and with TBF. The TBF are seen to act in total
{\it attractive} in the low density region, but turn to {\it
repulsive} at high densities, reflecting their specific density
dependence. The full D3Y interaction, including TBF, is used in
the following sections.

\section{Pairing at the Dripline}\label{sec:pair}

In nuclear matter pairing is found to be a low density phenomenon,
having a maximum contribution to the energy density around
$\frac{1}{3}$ of the saturation density $\rho_0=0.16 fm^{-3}$.
Therefore, it has to be expected that pairing is enhanced in
nuclei with a pronounced low-density region. This is exactly the
situation in 2- or multi-nucleon halo systems, and, to a lesser
extent, in weakly bound skin nuclei. A prominent case is $^{11}$Li
which exist as a particle-stable system mainly because of the
mutual interactions among the two valence neutrons. Their low
separation energy, S$_{2n}$=320~keV, points to the importance of
continuum coupling. For a detailed description of the valence wave
function the conventional BCS and HFB methods using
representations in terms of mean-field wave functions are not
suitable. Theoretically, continuum effects are properly described
by solving the Gorkov-equations \cite{go}
\begin{equation} \label{gorkov}
(h-e_{+})\Phi_{+} - \Delta \Phi_{-} = 0 \quad , \quad
(h-e_{-})\Phi_{-} + \Delta \Phi_{+} = 0
\end{equation}
for the hole and particle components \cite{le} $\Phi_\pm$, coupled
by the pairing field $\Delta$. The single particle energies are
$e_\pm=\lambda \pm E$ where $\lambda$ and E ($\geq 0$) denote the
chemical potential and the quasiparticle energy, respectively.
$\Phi_\pm$ will in general {\it not} coincide with the
eigenfunctions of the mean-field Hamiltonian $h$, i.e. the states
$\Phi_\pm$ are off the (mean-field) energy shell.

Particle-stability is given for $\lambda <0$. Then, irrespective
of the value of $E$, the hole-type solutions $\Phi_{-}$ are
exponentially decaying for $r \to \infty$. For E$\leq |\lambda|$,
an exponential asymptotic behaviour is also found for the
particle-type components $\Phi_+$ and a finite subset of discrete
eigenvalues $E$ is obtained. For E$> |\lambda|$, eq.(\ref{gorkov})
has to be solved with continuum wave boundary conditions for
$\Phi_{+}$ leading to single particle spectral functions being
distributed continuously in energy. Hence the quasi-particle
picture, underlying BCS theory and, to some degree, also
discretized HFB theory, is replaced by a fully dynamical continuum
description.

For like-particle $(S=0,T=1)$ pairing protons ($q=p$) and neutrons
($q=n$) are paired only among themselves. The pairing fields are
given in terms of the anomal or pairing density matrices $\kappa$
and the pairing interaction $V_{SE}$ \cite{Le01},
\begin{equation}
\Delta_q(r_1,r_2)=V_{SE}(r_1,r_2)\kappa_q(r_1,r_2) \quad .
\end{equation}
If $\lambda<0$, $\kappa_q(r)$ and $\Delta(r)$ are guaranteed to
decay exponentially for $r \to \infty$. It is still an open
question whether the free space or an in-medium singlet-even
interaction should be used (see e.g. \cite{discrete}).

\begin{figure}[t]
\begin{center}
\begin{minipage}[t]{8 cm}
\epsfig{file=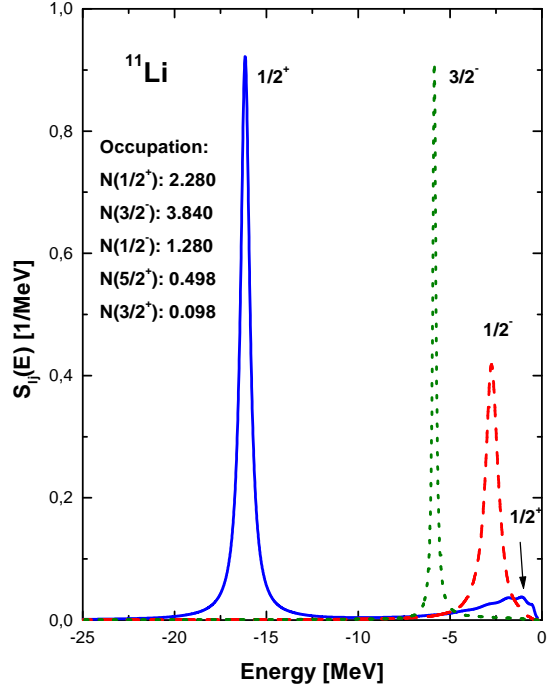,width=8cm}
\end{minipage}
\begin{minipage}[t]{12 cm}
\caption{Single particle spectral functions for s- and p-wave
neutron states in $^{11}$Li. The finite widths of the states are
due to continuum coupling. The partial occupation numbers
$N_{j\ell}$ are also shown, including d-wave contributions.
\label{fig3}}
\end{minipage}
\end{center}
\end{figure}

Strength functions for $^{11}$Li are displayed in Fig.\ref{fig3}.
The strong deviation from a pure mean-field or BCS description is
apparent by observing that besides the expected s- and p-wave
components also $d_{5/2,3/2}$ strength is lowered into the bound
state region. Remarkably, the mean-field does not support neither
bound $2s$ nor $1d_{5/2,3/2}$ single particle levels and their
appearance is solely due to the continuum coupling introduced by
pairing.

\begin{figure}[t]
\begin{center}
\begin{minipage}[t]{8 cm}
\epsfig{file=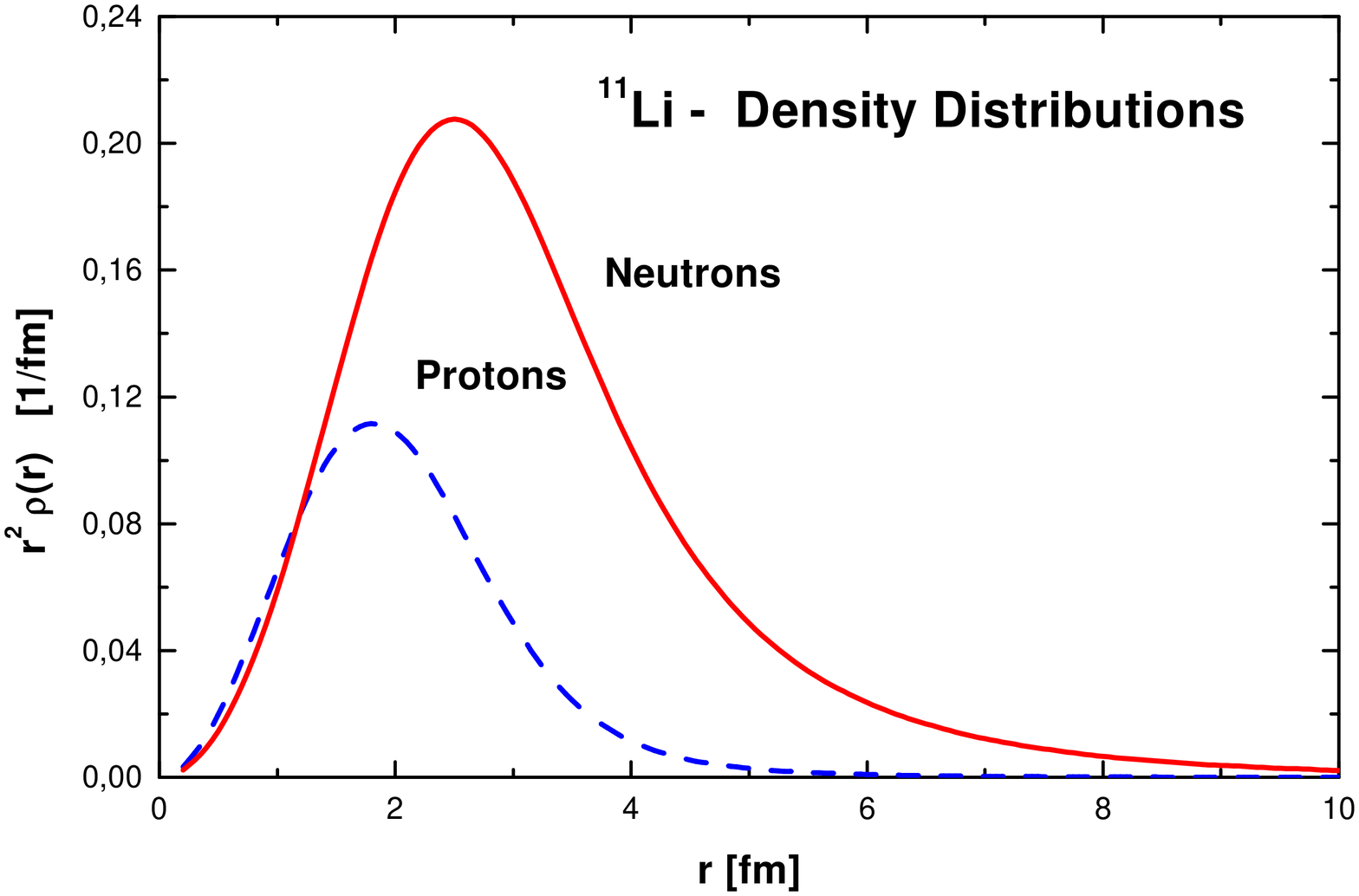,width=8cm}
\end{minipage}
\begin{minipage}[t]{12 cm}
\caption{Proton and neutron ground state density distributions
(weighted by $r^2$) for $^{11}$Li. The pronounced neutron halo
component formed by s-,p- and d-states is clearly visible.
\label{fig4}}
\end{minipage}
\end{center}
\end{figure}

Theoretically, the proton ($q=p$) and neutron ($q=n$) densities in
a systems like $^{11}$Li are defined by
\begin{equation}\label{density}
\rho_q(r)=\sum_{j\ell}{\frac{2j+1}{4\pi}\int^\lambda_{-\infty}{de_-
v^2_{qj\ell}(e_-)|G_{qj\ell}(r,e_-)|^2}} \quad ,
\end{equation}
where $\Phi_{-} = v^2 G$ was used \cite{Le01}. The particle
numbers are given by
\begin{equation}
N_q=\sum_{j\ell}{(2j+1)\int^\lambda_{-\infty}{de_-
v^2_{qj\ell}(e_-)}}=\sum_{j\ell}{(2j+1)n_{qj\ell}} \quad ,
\end{equation}
The neutron partial wave occupation numbers
$N_{j\ell}=(2j+1)n_{j\ell}$ for $^{11}$Li are displayed in
Fig.\ref{fig3}. In stable nuclei discrete levels at $2\lambda <
e_- < \lambda$ will contribute to eq.(\ref{density}). In extreme
dripline nuclei they are missing because of the smallness of
$|\lambda|$. In Fig.\ref{fig4} the proton and neutron ground state
densities are shown, multiplied by $r^2$, emphasizing the
differences in shape and the neutron halo component. High-energy
elastic scattering of $^{11}$Li on a proton target is well
described with the HFB densities \cite{egelh}. Measured $^{11}$Li
response functions were analyzed by QRPA calculations in
\cite{Zinser}.

\section{Dynamical Core Polarization at the
Dripline}\label{sec:dcp}

Dynamical core polarization (DCP) is seen most clearly in nuclei
with a single nucleon outside a core. Approaching the driplines
the core nucleus by itself is already far off stability with a
lower surface tension. Therefore, the restoring forces against
external perturbations are reduced. Such "soft core" systems are,
for example, found in the neutron-rich even-mass carbon isotopes.
A good indicator is the existence of low-energy 2$^+$ states,
decreasing in energy with increasing mass. Continuum QRPA
calculations with a residual interaction derived in Landau-Fermi
liquid theory from the D3Y in-medium interaction reproduce the
systematics of 2$^+$ states rather well. This leads to the
conclusion that they are mainly of vibrational nature rather than
due to static deformation as assumed e.g. in \cite{nun}.

The QRPA calculations predict a strong increase of the quadrupole
polarizabilities in the carbon isotopes for increasing neutron
excess with a maximum around $^{16,18}$C. This behaviour is due to
the lowering of the first 2$^+$ state which in $^{12}$C is located
at $E_x=4.44$~MeV and moves down to $E_x=1.66$~MeV in $^{18}$C.
The dipole (or electric) polarizability, however, changes only
within 10\% over the $^{10-22}$C isotopic chain.

\begin{figure}[t]
\begin{center}
\begin{minipage}[t]{8 cm}
\epsfig{file=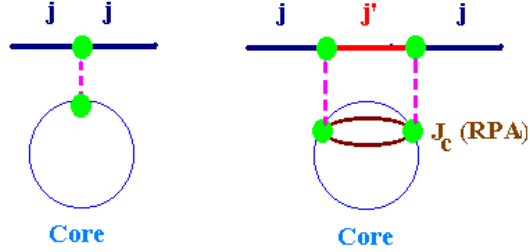,scale=0.7}
\end{minipage}
\begin{minipage}[t]{12 cm}
\caption{Diagramatic structure of mean-field (left) and core
polarization interactions (right) of a nucleon in single particle
state $j$. Interactions (meson exchange) are indicated by dashed
lines. Core polarization leads to intermediate states $(j'J_c)$
with a (QRPA) core excitation $J_c$ and a single particle state
$j'$. \label{fig5}}
\end{minipage}
\end{center}
\end{figure}

In a system with core polarization the valence particle (or hole)
obeys a non-static wave equation
\begin{equation}
\left ( H_{MF}+\Sigma_{pol}(\varepsilon)-\varepsilon \right
)\Psi=0
\end{equation}
including the static (HFB) mean-field Hamiltonian $H_{MF}$ and the
non-local and energy-dependent polarization self-energy
$\Sigma_{pol}$ describing the rescattering of the nucleon off the
core thereby exciting it into states of various multipolarities
and excitation energies followed by subsequent de-excitations back
to the ground state (see Fig.\ref{fig5}).

During these processes the particle can be scattered virtually
into high lying orbitals. The quantum numbers of the intermediate
$2p1h$ (or $1p2h$) configurations are only constraint by the
requirement that spin and parity must match those of the leading
particle configuration, given by the state moving with respect to
the core ground state. In nuclear matter and finite nuclei these
processes are known to give rise to a depletion of momentum space
ground state occupation probabilities \cite{dejong96,lehr00} from
the step function momentum distribution of a Fermi gas of
quasiparticles interacting only by a static mean-field. In
practice, the core excitations are calculated in QRPA theory thus
extending the static HFB picture in a consistent way to a
dynamical theory.

Dynamical single particle self-energies $\Sigma_{pol}$ affect
separation energies and wave functions. In dripline nuclei these
effects are strongly enhanced. In fact, dynamical self-energies
provide the main source of binding for $^{19}$C$(1/2^+,g.s.)$
\cite{Le01}.

Theoretically, the DCP description amounts to use a
multi-configuration ground state containing a single particle
component - reminiscent of the static mean-field configuration -
and a multitude of configurations where the valence particle is
rescattered into other orbits by interactions with the core
\cite{le,dcp}. Coupling to the lowest 2$^+$ and 3$^-$ states only
as advocated e.g. in \cite{vib} cannot account for the complexity
of the process.

\begin{table}[t]
\begin{center}
\begin{minipage}[t]{11 cm}
\caption{Energies, spins and ground state spectroscopic factors
from core polarization calculations. \label{tab1}}
\end{minipage}
\begin{tabular}{|c|c|c|c|}
\hline \raisebox{0pt}[13pt][7pt]{$Nucleus$}
&\raisebox{0pt}[13pt][7pt]{$j^\pi$} &
\raisebox{0pt}[13pt][7pt]{$Energy \; keV$} &\raisebox{0pt}[13pt][7pt]{$S(j^\pi,g.s.)$}\\
\hline
$^{8}B$   &$3/2^-$ & $130$ & $0.73$ \\
$^{11}Be$   &$1/2^+$ & $510$ & $0.74$ \\
$^{17}C$   &$5/2^+$ & $760$ & $0.43$ \\
$^{19}C$   &$1/2^+$ & $263$ & $0.41$ \\
\hline
\end{tabular}
\end{center}
\end{table}

The DCP wave functions have been used to analyze longitudinal
momentum distributions and one-nucleon removal cross sections in
relativistic breakup reactions with secondary beams
\cite{b8,c19,lola}, reproducing the data very satisfactorily.
Results for energies and spectroscopic factors in the single
neutron halo nuclei $^{11}$Be and $^{19}$C and the single proton
halo nucleus $^{8}$B \cite{b8} are shown in Tab.\ref{tab1}. In
$^{11}$Be core polarization is causing the reversal of $1/2^+$ and
$1/2^-$ states by supplying an additional attractive self-energy
in the 1/2$^+$ channel. The single particle spectroscopic factor
S$_n(1/2^*,g.s.)$=0.75 is in good agreement with recent transfer
and breakup data. A more dramatic effect is found in $^{19}$C and
also $^{17}$C where the g.s. components are strongly suppressed as
seen from the small spectroscopic factors in Tab.\ref{tab1}.
Hence, the valence nucleon is no longer attached to a definite
mean-field orbital but exists in a wave packet-like state spread
over a certain range of shell model states. In other words,
mean-field dynamics have ceased to be the dominant source of
binding. Rather, the $^{17}$C and $^{19}$C results indicate a new
type of binding mechanism in dripline nuclei where shell
structures are dissolved. The mean-field has lost its dominant
role and binding is produced by dynamical valence-core
interactions.

\section{Summary and Outlook}\label{sec:summary}

Asymmetric nuclear matter and nuclei far off stability were
described by microscopic approaches. An important advantage over
empirical models is the clean diagrammatic structure allowing to
extend the theory in a well defined way by inclusion of an
increasingly larger class of diagrams. This was illustrated on the
example of 3-body contributions to in-medium interactions. Pairing
and core polarization far off stability were discussed in some
detail. The results show that close to the dripline nuclear
dynamics change from mean-field to correlation dynamics. As an
important consequence shell structures are dissolved. This
reflects that the traditional understanding of nuclear structure
as being dominated by a static potential plus some residual
interactions cannot account for nuclear structure far off
stability.

\section*{Acknowledgments}
This work was supported in part by DFG, project Le 439/3, European
Graduate School "Complex Systems of Hadrons and Nuclei",
Giessen-Copenhagen, GSI Darmstadt and BMBF.

\end{document}